\documentclass[twocolumn,aps,showpacs,prb]{revtex4-1}

\usepackage{graphicx,epstopdf,amsmath,amssymb,multirow,CJK,color}
\usepackage[german,english]{babel}

\begin{document}

\title{First-principles study of the robust superconducting state of NbTi alloys under ultrahigh pressures}

\author{Jian-Feng Zhang$^{1}$}
\author{Miao Gao$^{2}$}
\author{Kai Liu$^{1}$}\email{kliu@ruc.edu.cn}
\author{Zhong-Yi Lu$^{1}$}\email{zlu@ruc.edu.cn}

\affiliation{$^{1}$Department of Physics and Beijing Key Laboratory of Opto-electronic Functional Materials $\&$ Micro-nano Devices, Renmin University of China, Beijing 100872, China}

\affiliation{$^{2}$Department of Physics, School of Physical Science and Technology, Ningbo University, Zhejiang 315211, China}

\date{\today}

\begin{abstract}

A recent experiment reported that robust superconductivity appears in NbTi alloys under ultrahigh pressures with an almost constant superconducting $T_c$ of $\sim$19 K from 120 to 261.7 GPa [J. Guo \textit{et al.}, Adv. Mater. {\bf 31}, 1807240 (2019)], which is very rare among the known superconductors. We investigate the origin of this novel superconducting behavior in NbTi alloys based on density functional theory and density functional perturbation theory calculations. Our results indicate that the pressure tends to transform NbTi alloys from a random phase to a uniformly ordered crystal phase, and the exotic robust superconductivity of NbTi alloys can still be understood in the framework of BCS theory. The Nb element in NbTi alloys plays a dominant role in the superconductivity at low pressure, while the NbTi crystal with an alternative and uniform Nb and Ti atomic arrangement may be responsible for the stable superconductivity under high pressures. The robust superconducting transition temperature of NbTi under ultrahigh pressure can be explained by a synergistic effect of the enhanced phonon frequency, the modestly reduced total electron-phonon coupling, and the pressure-dependent screened Coulomb repulsion.

\end{abstract}

\date{\today} \maketitle

%\pacs{}

\maketitle

\section{INTRODUCTION}

Pressure has played an important role in both synthesizing new superconductors and modulating the superconductivity of existing materials~\cite{RMP2019}.
For unconventional superconductors, the highest superconducting transition temperature $T_c$ of 164 K was achieved in the cuprate HgBa$_2$Ca$_2$Cu$_3$O$_{8+\delta}$ under 31 GPa~\cite{cwc94}; in comparison, the highest $T_c$ of iron-based superconductors was observed in rare-earth iron oxyarsenides~\cite{Zhi-An Ren, Zhu-An Xu}, while SmFeAsO$_{1-x}$F$_x$ was synthesized under 6 GPa~\cite{Zhi-An Ren}.
On the side of conventional superconductors, according to BCS theory, dense hydrogen was estimated to become a high-$T_c$ superconducting metal at high pressure half a century ago~\cite{Aschcroft}. Later, hydrides were prepared under ultrahigh pressures to avoid the difficulty in metallizing hydrogen~\cite{Drozdov15,lah1,lah2}. Several experimental breakthroughs have taken place in recent years: sulfur hydride shows a $T_c$ of 203 K at 155 GPa~\cite{Drozdov15} and lanthanum superhydride possesses a record $T_c$ of $\sim$260 K around 190 GPa~\cite{lah1,lah2}, while the prominent isotope effect indicates their conventional characters~\cite{Drozdov15,lah1,lah2}.
Beyond those superconducting compounds consisting of metallic and nonmetallic elements, the superconducting alloys are also a large family of superconductors. Recently, robust superconductivity was observed in several superconducting alloys from ambient to ultrahigh pressures~\cite{cavapnas,sllpnas,sllarxiv,sunll19,cavaprm,sunll20}.
%Nevertheless, the study on the underlying mechanism of this exotic superconducting phenomenon is just beginning~\cite{sunll20}.

Among the alloy superconductors, NbTi alloys are a well-known commercial superconducting material with superior properties, such as high critical magnetic field $H_{c2}$, high critical current density $I_c$, an accessible superconducting transition temperature $T_c$, easy workability, etc.~\cite{nbtiapp1,nbtiapp2,sllarxiv}. A fresh experiment reported the robust superconductivity in NbTi alloys under ultrahigh pressures~\cite{sllarxiv}, where $T_c$ of 19 K and $H_{c2}$ of 19 T set the corresponding records among all known transition-metal-alloy superconductors. Astonishingly, NbTi alloys remain at an almost constant $T_c$ of $\sim$19 K in a wide pressure range from 120 to 261.7 GPa and keep the bcc lattice structure under all measured pressures~\cite{sllarxiv}. Similar exceptional behavior has been observed in high-entropy-alloy~\cite{aem2004,sunll19} superconductors such as the Nb-Ta-Zr-Hf-Ti alloy~\cite{cavapnas,sllpnas,cavaprm,jasiewicz,sunll19} (see Ref. \onlinecite{sunll19} for a recent review). These phenomena are in stark contrast to the sensitive pressure dependence of $T_c$ for most superconductors. A recent work suggested that the electrons from $e_g$ orbitals are likely responsible for the stable $T_c$~\cite{sunll20}. Nevertheless, several interesting questions need to be explored. Is the electron-phonon coupling (EPC) mechanism of the BCS theory still applicable to NbTi alloys at ultrahigh pressure, or is there some other novel mechanism? Why is the superconductivity so robust under such a wide pressure range?

We have investigated the origin of this exotic superconducting behavior of the model superconducting alloy NbTi. By performing systematic first-principles calculations on the evolution of the electronic structure and phonon spectrum of NbTi with pressure, we show that the pressure tends to drive the NbTi alloy from a random atomic arrangement to a CsCl-type structure with alternative Nb and Ti distributions (NbTi crystal); meanwhile, the bcc lattice structure remains. To understand the robust superconductivity observed in experiment~\cite{sllarxiv}, we further calculated the superconducting $T_c$ of the NbTi crystal in the pressure range from 50 to 250 GPa based on EPC theory.

\section{Method}

The electronic structure, phonon dispersion, and EPC strength of the NbTi crystal under the pressure range from 0 to 250 GPa were studied with density functional theory \cite{dft1,dft2} and density functional perturbation theory \cite{dfpt,dfptreview} calculations as implemented in the QUANTUM ESPRESSO (QE) package \cite{pwscf}. The interactions between electrons and nuclei were described by the norm-conserving pseudopotentials \cite{ncpp}. The valence electron configurations were $4s^2 4p^6 4d^4 5s^1$ for Nb and $3s^2 3p^6 3d^2 4s^2$ for Ti. For the exchange-correlation functional, the generalized gradient approximation of Perdew-Burke-Ernzerhof \cite{PBE} type was adopted. The kinetic energy cutoff of the plane-wave basis was set to be 160 Ry. The Gaussian smearing method with a width of 0.004 Ry was employed for the Fermi surface broadening. In structural optimization, both lattice constants and internal atomic positions were fully relaxed until the forces on all atoms were smaller than 0.0002 Ry/bohr.

The superconducting $T_c$ of the NbTi crystal was calculated based on the EPC theory as implemented in the EPW package \cite{epw}, which uses the maximally localized Wannier functions \cite{mlwf} and interfaces with QE \cite{pwscf}. We took the 6$\times$6$\times$6 {\bf k} mesh and {\bf q} mesh as coarse grids and interpolated to the 108$\times$108$\times$108 {\bf k}-mesh and 12$\times$12$\times$12 {\bf q}-mesh dense grids. The EPC constant $\lambda$ can be calculated either by the summation of the EPC constant $\lambda_{{\bf q}\nu}$ in the full Brillouin zone for all phonon modes or by the integral of the Eliashberg spectral function $\alpha^2F(\omega)$ as
\begin{equation}
\lambda=\sum_{{\bf q}\nu}\lambda_{{\bf q}\nu}=2\int{\frac{\alpha^2F(\omega)}{\omega}d\omega}.
\end{equation}
The Eliashberg spectral function $\alpha^2F(\omega)$ is defined as\cite{Eliashberg}
\begin{equation}
\alpha^2F(\omega)=\frac{1}{2{\pi}N(\varepsilon_F)}\sum_{{\bf q}\nu}\delta(\omega-\omega_{{\bf q}\nu})\frac{\gamma_{{\bf q}\nu}}{\hbar\omega_{{\bf q}\nu}},
\end{equation}
where $N(\varepsilon_F)$ is the density of states (DOS) at the Fermi level $\varepsilon_F$, $\omega_{{\bf q}\nu}$ is the frequency of the $\nu$th phonon mode at the wave vector {\bf q}, and $\gamma_{{\bf q}\nu}$ is the phonon linewidth.

The superconducting transition temperature $T_c$ can be estimated by substituting the EPC constant $\lambda$ into the McMillan-Allen-Dynes formula \cite{mcmillan1, mcmillan2},
\begin{equation}
{k_B}T_c=\frac{\hbar\omega_{log}}{1.2}\text{exp}[\frac{-1.04(1+\lambda)}{\lambda(1-0.62\mu^*)-\mu^*}],
\end{equation}
where $\mu^*$ is an effective screened Coulomb repulsion constant. In general, $\mu^*$ is treated as a semiempirical parameter\cite{dfptreview} between 0.1 and 0.2. In our calculations, $\mu^*$ was set by two strategies for comparison: (1) a constant\cite{dfptreview} of 0.15 and (2) a pressure-dependent one\cite{aip2008} with $\mu^*=0.26N(\varepsilon_F)/[1+N(\varepsilon_F)]$. $\omega_{log}$ is the logarithmic average of the Eliashberg spectral function,
\begin{equation}
\omega_{log}=\text{exp}[\frac{2}{\lambda}\int{\frac{d\omega}{\omega}\alpha^2F(\omega){\text{ln}}(\omega)}].
\end{equation}

\begin{figure}[tb]
\includegraphics[angle=0,scale=0.3]{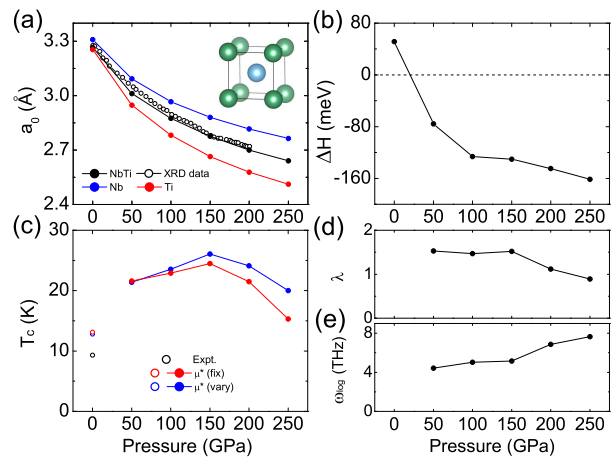}
\caption{(Color online) (a) Calculated lattice constants of the NbTi crystal (black dots), Nb elemental crystal (blue dots), and Ti elemental crystal (red dots) in comparison with the measured values of NbTi alloys in a previous XRD experiment \cite{sllarxiv} (black circles) under pressure. The inset shows the structure of CsCl-type NbTi crystal. (b) Enthalpy differences $\Delta H$ between the NbTi crystal and the Nb and Ti elemental crystals: $\Delta H$ = $H(\text{NbTi}) - [H(\text{Nb}) + H(\text{Ti})]$. Calculated (c) superconducting $T_c$, (d) total EPC $\lambda$, and (e) $\omega_{log}$ of the NbTi crystal in the pressure range of 50 to 250 GPa. In (c), the red and blue circles represent the calculated $T_c$ of the NbTi crystal with $\mu^*$ set to 0.15 and $0.26N(\varepsilon_F)/[1+N(\varepsilon_F)]$, respectively. The open circles represent the calculated (red and blue) and experimental (black) superconducting $T_c$ of Nb elemental crystal at ambient pressure.}
\label{fig1}
\end{figure}

\section{Results}

According to a previous x-ray diffraction (XRD) experiment \cite{sllarxiv}, the NbTi alloy retains a bcc structure under all measured pressures. In real NbTi alloys, the Nb and Ti atoms may distribute randomly in a bcc lattice. We thus considered two limits: one is the uniform ``NbTi crystal" with the Nb and Ti atoms arranging alternately [as shown in the inset of Fig. 1(a)]; the other is the pure Nb and Ti elemental crystals that also have a bcc structure but locate in different domains of materials. A real NbTi alloy is, to some degree, between these two extreme forms. Figure 1(a) shows the calculated lattice constants of the NbTi \textit{crystal} (black dots) along with the measured values of the NbTi \textit{alloy} (black circles) with increasing pressure\cite{sllarxiv}, where good agreement between them can be observed. For comparison, the evolutions of calculated lattice constants of Nb (blue dots) and Ti (red dots) elemental crystals are also demonstrated, which both show increasing deviations from the experimental results under pressure. Figure 1(b) displays the calculated enthalpy differences $\Delta H$ between the NbTi crystal and the pure Nb and Ti elemental crystals in the same pressure range. A sign of transition takes place around 20 GPa, indicating that high pressure tends to drive the NbTi alloy to the uniform NbTi crystal. Furthermore, we also calculated the enthalpies of special NbTi alloys with alternating Nb and Ti layers, which are larger ($>$0.04 eV/atom at 100 GPa) than that of uniform NbTi crystal. Nevertheless, the transition barriers from those special NbTi alloys to uniform NbTi crystal are less than 0.2 eV/atom at 100 GPa, lower than the one between graphite and diamond under pressure\cite{graph2dia}. This suggests the high possibility of the occurrence of this transition (see Appendix A for more details). Therefore, employing NbTi crystal to simulate NbTi alloy under high pressure is reasonable.

\begin{figure}[t]
\includegraphics[angle=0,scale=0.33]{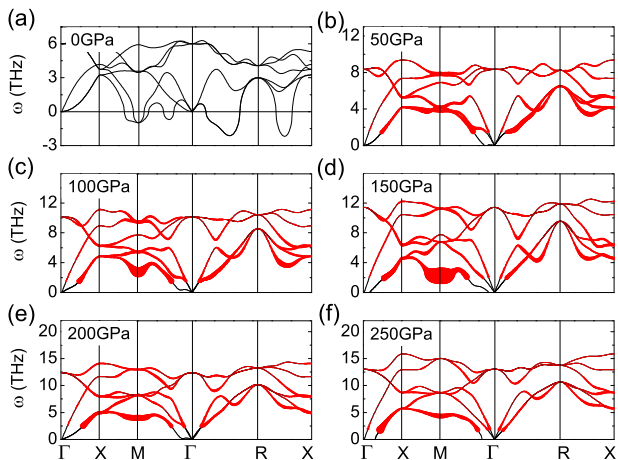}
\caption{(Color online) Phonon dispersions of the NbTi crystal in the pressure range of 0 to 250 GPa at an interval of 50 GPa. The sizes of the red dots on the phonon dispersions are proportional to the EPC strengths $\lambda_{{\bf q}\nu}$. The low-frequency region was cut off to exclude the influence of error around the $\Gamma$ point.}
\label{fig2}
\end{figure}

To investigate the dynamical stability of NbTi crystal, we calculated its phonon dispersions under different pressures (Fig. 2). At 0 GPa, there is a large portion of imaginary frequencies across the Brillouin zone: the $M$ point, the $\Gamma$-$M$ path, the $\Gamma$-$R$ path, the $R$-$X$ path, and so on. This indicates that the uniform NbTi crystal is unstable at ambient pressure, which is also in accordance with the above enthalpy difference between the NbTi crystal and the Nb/Ti elemental crystals [Fig.1(b)]. The corresponding structural distortions of NbTi may be quite complex, implying the chaotic phase of NbTi alloys and random distributions of Nb/Ti atoms at 0 GPa.
In fact, similar destabilization of the bcc phase at low pressure was also reported for Nb metals alloyed with Ti\cite{vitosprb}. The phonon dispersions of the NbTi crystal in the pressure range from 50 to 250 GPa at an interval of 50 GPa are shown in Figs. 2(b) to 2(f). In these cases, almost all imaginary frequencies disappear except for some tiny ones around the $\Gamma$ point, suggesting that the NbTi crystal is dynamically stable under high pressures, which agrees with the lower enthalpies of the NbTi crystal compared with Nb/Ti elemental crystals beyond 20 GPa [Fig. 1(b)]. With increasing pressure, the phonon modes around the $M$ point show dramatic changes; meanwhile, the topmost phonon modes shift to higher frequencies due to the strengthened bonding between Nb and Ti atoms.

\begin{table}[!b]
\caption{The calculated electronic density of states at the Fermi level $N(\varepsilon_F)$ [in units of states/(eV atom)] and effective screened Coulomb repulsion constant $\mu^*$ based on the relation: $\mu^*=0.26N(\varepsilon_F)/[1+N(\varepsilon_F)]$ under different pressures $P$ (in GPa) for the NbTi crystal. The values at 0 GPa are from Nb elementary crystal.}
\begin{center}
\begin{tabular*}{8cm}{@{\extracolsep{\fill}} ccccccc}
\hline \hline
$P$ & 0 & 50 & 100 & 150 & 200 & 250  \\
\hline
$N(\varepsilon_F)$ & 1.49 & 1.42 & 1.20 & 1.02 & 0.94 & 0.72  \\
$\mu^*$ & 0.155 & 0.153 & 0.142 & 0.131 & 0.126 & 0.108  \\
\hline\hline
\end{tabular*}
\end{center}
\end{table}

We further studied the superconducting properties of the NbTi crystal based on the EPC theory \cite{Eliashberg} and calculated the superconducting transition temperature $T_c$ according to the McMillan-Allen-Dynes formula [Eq. (3)] \cite{mcmillan1,mcmillan2}. Here two strategies were adopted for the effective screened Coulomb repulsion constant $\mu^*$. The first approach uses a constant $\mu^*$ of 0.15 [red circles in Fig. 1(c)]. As can be seen, the superconductivity in the NbTi crystal is very robust against pressure. $T_c$ increases from 21.6 K at 50 GPa to 24.5 K at 150 GPa and then remains above 20 K until 200 GPa, after which it begins to drop gradually. The second approach sets $\mu^*$ as $0.26N(\varepsilon_F)/[1+N(\varepsilon_F)]$. Detailed values of $N(\varepsilon_F)$ and $\mu^*$ at different pressures are shown in Table I. With the pressure-dependent $\mu^*$, the dropping of $T_c$ after 150 GPa weakens, and $T_c$ appears more robust under high pressure [blue circles in Fig. 1(c)]. To further confirm the reliability of our calculations, we also calculated the superconducting $T_c$ of Nb elemental crystal. Its stable $T_c$ at low pressures and rapid dropping at high pressures\cite{nbpre} were also reproduced (see Fig. S1(a) in the Supplemental Material \cite{suppmat}).

The pressure-dependent total EPC constant $\lambda$ and logarithmic frequency $\omega_{log}$ of the NbTi crystal are summarized in Figs. 1(d) and 1(e), respectively. As the pressure increases, $\lambda$ remains almost stable up to 150 GPa and then decreases gradually. Meanwhile, $\omega_{log}$ demonstrates a slow growth below 150 GPa and then increases with pressure. These can be understood via the variations of phonon dispersions. As can be seen in Fig. 2, the pressure hardens the phonon frequency on the whole. Nevertheless, the frequency of the lowest acoustic mode at the $M$ point, i.e., the $A_{2u}$ mode that provides the largest $\lambda_{{\bf q}\nu}$ [as indicated by the biggest red dots in Fig. 2(c)], shows a nonmonotonic variation: it first decreases and then increases with pressure. Between 50 and 150 GPa, the initial frequency decrease of the $A_{2u}$ mode offsets the partial influence of pressure on all phonon spectra, resulting in a slowly growing $\omega_{log}$. %The abnormal soften of this $A_{2u}$ mode maybe related to the topography of Fermi surface. But unfortunately their relations aren't distinct, as discussed in Appendix C.
Furthermore, there are also contributions from other phonon modes to the EPC $\lambda_{{\bf q}\nu}$, which are proportional to the sizes of red dots on the phonon dispersions in Figs. 2(b)-2(f). %\textit{Explain the stable $\lambda$ and $T_c$.}

%\begin{figure}[tb]
%\includegraphics[angle=0,scale=0.25]{Fig2.eps}
%\caption{(Color online) Calculated (a) superconducting $T_c$, (b) total EPC $\lambda$, and (c) $\omega_{log}$ of NbTi crystal under different pressures (50 - 250 GPa). In panel (a), the red and blue circles represent the calculated $T_c$ with $\mu^*$ setting as 0.15 and $0.26N(\varepsilon_F)/[1+N(\varepsilon_F)]$, respectively. The hollow circles represent the calculated (red/blue) and experimental (black) superconducting $T_c$ of Nb elementary crystal at ambient pressure.}
%\label{fig3}
%\end{figure}

\section{Discussion and Summary}

\begin{figure}[tb]
\includegraphics[angle=0,scale=0.32]{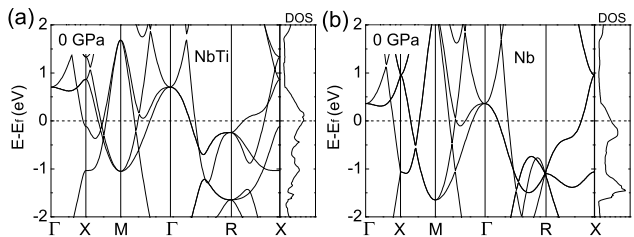}
\caption{(Color online) Band structures and density of states (DOS) of (a) the NbTi crystal and (b) Nb elemental crystal at 0 GPa. For comparison convenience, the band structure and DOS of Nb elemental crystal were calculated in a cubic cell that contains two Nb atoms.}
\label{fig3}
\end{figure}

Both the calculated enthalpies [Fig. 1(b)] and the lattice dynamics (Fig. 2) indicate that the pressure tends to drive the NbTi alloy to be uniformly ordered. According to the fundamental thermodynamics formula, $G=H-TS$, where $G$ is the Gibbs free energy, $H$ is the enthalpy, $T$ is the temperature, and $S$ is the entropy. In the synthesis process at high temperature and ambient pressure, the large entropy $S$ (degree of disorder) is conducive to minimizing the free energy and to stabilizing the system. Nevertheless, at low temperature and high pressure, the contribution of entropy $S$ to the free energy is much reduced and the enthalpy $H$ dominates the free energy. As a result, the NbTi crystal with a lower enthalpy [Fig. 1(b)] and bond lengths closer to the experimental ones [Fig. 1(a)] compared with those of Nb and Ti elements is energetically favored under high pressures, which makes the NbTi alloy more uniform.

To analyze the origin of higher $T_c$ of the NbTi crystal than that of the Nb elemental crystal, we compare their band structures and DOSs at the 0 GPa (Fig. 3). First, similar to Nb elemental crystal with the highest superconducting $T_c$ (9.3 K) among all elements, the NbTi crystal also possesses a bcc lattice structure, and so do many other Nb-based superconducting alloys~\cite{cavapnas,sllpnas}. From Fig. 3, we can see that the band structures of the NbTi crystal and Nb element are quite similar except for some band splitting due to the lower symmetry of the crystal field (cubic) in the former one compared with the latter one (bcc). Second, since the Ti atom has one less valence electron than the Nb atom, the NbTi crystal has a lower Fermi level, which just locates on a peak in the DOS [Fig. 3(a)]. Based on the BCS theory \cite{bcs}, a large DOS at the Fermi level $N(\varepsilon_F)$ is beneficial for the appearance of superconductivity. In previous studies on Nb-based high-entropy alloys, it was reported that the superconductivity is in close relation to the average valence electron $n$ and the optimal $T_c$ appears around $n=4.67$, which corresponds to 0.33 hole doping per Nb atom~\cite{cavapnas,cavaprm,sunll19}. Here the hole doping induced by the Ti atom in NbTi alloys is thus helpful to boost the superconductivity. Last but not least, Ti has a much lighter atomic mass than Nb, which can provide higher phonon frequencies. According to the McMillan-Allen-Dynes formula [Eq. (3)], $T_c$ is proportional to the logarithmic average phonon frequency $\omega_{log}$ [Eq. (4)]. So the higher phonon frequency introduced by the Ti atom is also conducive to enhancing the superconducting $T_c$. In fact, a compound composed of light elements has long been pursued to realize high superconducting $T_c$, such as MgB$_2$ \cite{mgb2}, H$_3$S \cite{Drozdov15}, and lanthanum superhydride \cite{lah1,lah2,lah3}. Thus, the hole doping and higher phonon frequency induced by the Ti atom are crucial for the higher $T_c$ of the NbTi crystal compared with the Nb element.

The robust high-$T_c$ superconductivity in NbTi alloys under high pressures~\cite{sllarxiv} can be understood as follows. At ambient pressure, the NbTi alloy is in such a phase that the Nb and Ti atoms distribute in random domains. It turns out that Nb atoms may maintain their elemental superconducting property. Actually, the observed superconducting $T_c$ of 9.6 K in NbTi alloys at ambient pressure\cite{sunll19} is very close to that of the Nb element (9.3 K) [Fig. 1(c)]~\cite{nbpre}. Under high pressure, the NbTi alloy gradually transforms to a uniform NbTi crystal phase [Fig. 1(b)], as verified by the consistent lattice constants between our calculations and the measured ones [Fig. 1(a)]. The homogeneous arrangement of Nb/Ti atoms in the NbTi crystal under high pressure promotes the electron transfer from Nb to Ti, which increases the density of states at the Fermi level (Fig. 3) and thus enhances the superconducting $T_c$. In the pressure range from 50 to 150 GPa, the $A_{2u}$ phonon mode at the $M$ point with strong EPC has an abnormal softening (Fig. 2). This offsets partial influences of pressure on $\lambda$ and $\omega_{log}$ [Figs. 1(d) and 1(e)], thus reducing the variation of $T_c$. At 150 GPa, the phonon softening at the $M$ point becomes quite strong. Its relation to the electronic structure is discussed in Appendix B.
Above 150 GPa, our calculated superconducting $T_c$ shows a drop compared with the observed robust $T_c$ until 261 GPa~\cite{sllarxiv}. This can be explained from two sides. On the one hand, NbTi alloys cannot completely transform to a single NbTi crystal. Although the ultrahigh pressure energetically tends to make more Nb and Ti atoms arrange alternatively and uniformly, there is still a portion of atoms in the disordered distribution. As a result, there may be a hysteresis behavior for the evolution of superconductivity with pressure.
%, which is reflected in the fact that the experimental $T_c$ arrives the optimal value of 19 K at 150 GPa but our calculated $T_c$ of NbTi crystal already exceeds 20 K at 50 GPa [Fig. 3(a)].
On the other hand, the pressure-dependent $\mu^*$ can bring more robust $T_c$ than the constant $\mu^*$ [Fig. 1(c)]. According to the classic relation \cite{aip2008} $\mu^*=0.26N(\varepsilon_F)/[1+N(\varepsilon_F)]$, $\mu^*$ has a positive correlation with the DOS at the Fermi level. Compared with the value at 50 GPa, $N(\varepsilon_F)$ is reduced by about 50\% at 250 GPa (Table I), which leads to a smaller $\mu^*$. According to the McMillan-Allen-Dynes formula [Eq. (3)], the decreased $\lambda$ [Fig. 1(d)], increased $\omega_{log}$ [Fig. 2(c)], and reduced $\mu^*$ under high pressure will compensate each other and result in stable $T_c$ of the NbTi crystal.

Our studies reveal that the pressure tends to transform NbTi alloys from a random phase to a uniformly ordered crystal phase (NbTi crystal). Therefore, we suggest that at ambient pressure the superconductivity mainly comes from the Nb elements in the alloy, while under high pressures it is mainly contributed by the NbTi crystal. The calculated superconducting $T_c$ of the uniform NbTi crystal with pressure-dependent $\mu^*$ can remain above 20 K until 250 GPa, where the softened phonon mode at the $M$ point with the largest electron-phonon coupling counteracts the partial influence of pressure. This gives a reasonable explanation for the observed robust superconductivity of NbTi alloys under high pressure~\cite{sllarxiv} and indicates that the EPC mechanism of BCS theory is still applicable to NbTi under extreme conditions. In addition to NbTi alloys, our studies may also apply to other alloy superconductors, which provides insight into tuning $T_c$ of superconducting alloys by controlling their atomic randomness in synthetic process.

\begin{acknowledgments}

We thank L.-L. Sun for providing the experimental lattice data and are grateful to L.-L. Sun, Q. Wu, and T. Xiang for helpful discussions. This work was supported by the National Key R$\&$D Program of China (Grants No. 2017YFA0302903 and No. 2019YFA0308603), the National Natural Science Foundation of China (Grants No. 11774422, No. 11774424, and No. 11974194), the CAS Interdisciplinary Innovation Team, the Fundamental Research Funds for the Central Universities, and the Research Funds of Renmin University of China (Grant No. 19XNLG13). Computational resources were provided by the Physical Laboratory of High Performance Computing at Renmin University of China.

\end{acknowledgments}

\begin{appendix}

\section{Transition barriers of NbTi alloys}

To confirm the possibility of transformation from a disordered atomic arrangement to a uniform one in NbTi alloys, we propose a transition process exemplified by the one from a NbTi alloy with alternating Nb and Ti layers along the [110] direction [Fig. 4(a), space group $Cmmm$] to the uniform CsCl-type NbTi crystal [Fig. 4(c), space group $Pm-3m$]. This process undergoes a metastable state with Nb and Ti bilayers arranged alternatively along the [112] direction [Fig. 4(b), space group $P2_1/m$]. Note that all three structures possess a bcc lattice but different atomic arrangements. The calculated transition barriers of the above transition process are shown in Fig. 4(d). At 100 GPa, the transition barriers are less than 0.2 eV/atom, which is lower than that between graphite and diamond under pressure (0.22 eV/atom)\cite{graph2dia}.

\begin{figure}[tb]
\includegraphics[angle=0,scale=0.31]{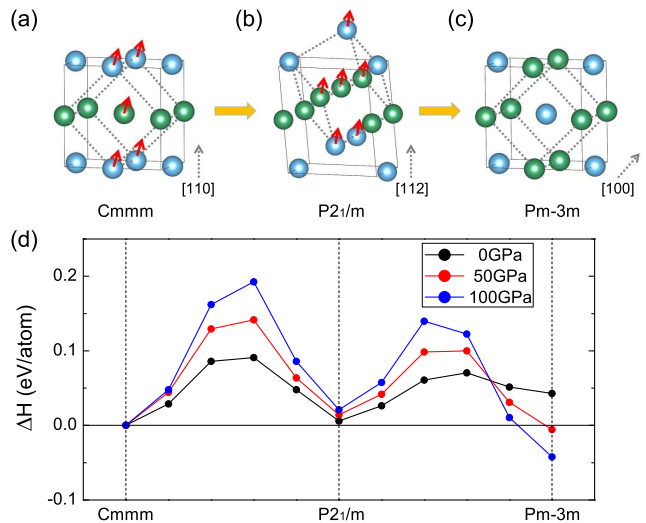}
\caption{(Color online) (a)-(c) Proposed transition process exemplified by the one from a NbTi alloy with alternating Nb and Ti layers along the [110] direction (space group $Cmmm$) to the uniform CsCl-type NbTi crystal (space group $Pm-3m$), which undergoes a metastable state with Nb and Ti bilayers arranged alternatively along the [112] direction (space group $P2_1/m$). (d) Calculated barriers of the above transition process.}
\label{s1}
\end{figure}

\begin{figure}[tb]
\includegraphics[angle=0,scale=0.3]{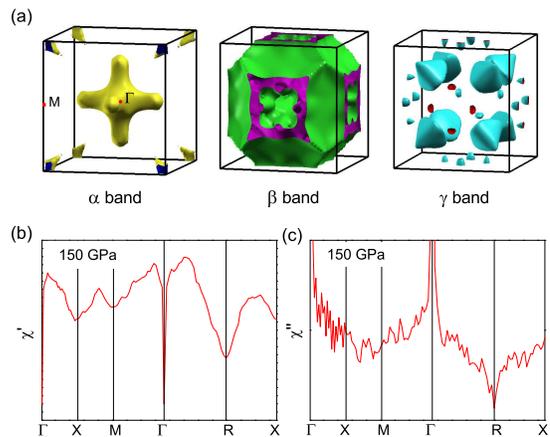}
\caption{(Color online) (a) Calculated Fermi surface sheets of the NbTi crystal at 150 GPa. (b) Real and (c) imaginary parts of electronic susceptibility $\chi$ of the NbTi crystal at 150 GPa}
\label{s2}
\end{figure}

\section{Relation between electronic structure, phonon dispersion, and electron-phonon coupling}

The strong electron-phonon coupling usually induces an abnormal phonon softening, as demonstrated in the calculated phonon dispersions of the NbTi crystal [Figs. 2(b)-2(f)]. Recalling the formula of the phonon linewidth,
\begin{equation}
\gamma_{{\bf q}\nu}=2\pi\omega_{{\bf q}\nu}\sum_{{\bf k}nn'}|g_{{\bf k+q}n',{\bf k}n}^{{\bf q}\nu}|^2\delta(\varepsilon_{{\bf k}n}-\varepsilon_F)\delta(\varepsilon_{{\bf k+q}n'}-\varepsilon_F),
\end{equation}
where $g_{{\bf k+q}n',{\bf k}n}^{{\bf q}\nu}$ is the EPC matrix element and $\sum_{{\bf k}nn'}{\delta(\varepsilon_{{\bf k}n}-\varepsilon_F)\delta(\varepsilon_{{\bf k+q}n'}-\varepsilon_F)}$ is the function of Fermi-surface nesting, perfect nesting of the Fermi surface along the $\Gamma$-$M$ vector may give some information about the phonon softening around the $M$ point. We thus calculated the Fermi surface of the NbTi crystal under 150 GPa [Fig. 5(a)], at which pressure NbTi has the most obvious phonon softening at the $M$ point [Fig. 2(d)]. As shown in Fig. 5(a), the Fermi surfaces are very complex, and it is hard to make a direct connection with phonon softening. For better visualization, we further calculated the electronic susceptibility $\chi$, whose real and imaginary parts are defined as: $\chi'({\bf q})=\sum_{nn'{\bf k}}{\frac{f({\varepsilon_{{\bf k}n})-f(\varepsilon_{{\bf k+q}n'})}}{\varepsilon_{{\bf k}n}-\varepsilon_{{\bf k+q}n'}}}$ and $\chi''({\bf q})=\sum_{nn'{\bf k}}{\delta(\varepsilon_{{\bf k}n}-\varepsilon_{F})\delta(\varepsilon_{{\bf k+q}n'}-\varepsilon_{F})}$, respectively. According to Figs. 5(b) and 5(c), neither $\chi'$ nor $\chi''$ shows a peak at the $M$ point. These results indicate that the pure electronic structure of NbTi (Fig. 5) cannot directly reflect the variations of EPC and phonon softening (Fig. 2).

\end{appendix}

\end{document}